\documentclass[pdflatex,sn-mathphys-num]{sn-jnl}

\usepackage{graphicx}%
\usepackage{multirow}%
\usepackage{amsmath,amssymb,amsfonts}%
\usepackage{amsthm}%
\usepackage{mathrsfs}%
\usepackage[title]{appendix}%
\usepackage{xcolor}%
\usepackage{textcomp}%
\usepackage{manyfoot}%
\usepackage{booktabs}%
\usepackage{algorithm}%
\usepackage{algorithmicx}%
\usepackage{algpseudocode}%
\usepackage{listings}%
\usepackage{ulem}

\theoremstyle{thmstyleone}%
\theoremstyle{thmstyletwo}%

\theoremstyle{thmstylethree}%

\begin{document}

\title[Article Title]{Modular fabrication and design of thick rigid-foldable origami metamaterials}


\author*[1]{\fnm{Sunao} \sur{Tomita}}\email{stomita@mosk.tytlabs.co.jp}
\author[1]{\fnm{Hiroki} \sur{Kobayashi}}\email{hiroki.kobayashi@mosk.tytlabs.co.jp}
\author[1]{\fnm{Shoko} \sur{Arita}}\email{shoko.arita.fs@mosk.tytlabs.co.jp}
\author[1]{\fnm{Masato} \sur{Tanaka}}\email{tanamasa@mosk.tytlabs.co.jp}
\author[1]{\fnm{Atsushi} \sur{Kawamoto}}\email{atskwmt@mosk.tytlabs.co.jp}
\author[1]{\fnm{Tsuyoshi} \sur{Nomura}}\email{nomu2@mosk.tytlabs.co.jp}

\author*[2]{\fnm{Tomohiro} \sur{Tachi}}\email{tachi@idea.c.u-tokyo.ac.jp}



\affil*[1]{\orgname{Toyota Central R\&D Labs., Inc.}, \orgaddress{\street{1-4-14 
Koraku}, \city{Bunkyo-ku}, \postcode{112-0004}, \state{Tokyo}, \country{Japan}}}

\affil*[2]{\orgdiv{Department of General Systems Studies, Graduate School of Arts and Sciences}, \orgname{The University of Tokyo}, \orgaddress{\street{3-8-1 Komaba}, \city{Meguro-ku}, \postcode{153-8902}, \state{Tokyo}, \country{Japan}}}


\abstract{
Origami metamaterials offer significant potential for stiff deployable structures.
However, fabricating load-bearing cellular structures from thick panels introduces geometric interference at non-manifold junctions.
Conventional thick-panel fabrication often disrupt ideal kinematics, thereby compromising smooth motion and scalability.
This study proposes a modular fabrication framework that preserves one-degree-of-freedom rigid-folding kinematics in thick and non-manifold origami metamaterials.
By decomposing non-manifold junctions into a hierarchy of stacked, modular hinged panels, our approach successfully accommodates synchronized hinge motions using scissor-like linkages.
Exploiting this representation, we implement a graph-based topology optimization framework that tailors macroscopic stiffness while preserving folding connectivity.
We demonstrate this approach by fabricating optimized prototypes that deploy seamlessly with a one-degree-of-freedom motion.
Furthermore, we demonstrate engineering scalability through the large-scale construction of extensive deployable systems assembled from modular panels, which exhibit high load-bearing capacity.
These results pave the way for the practical fabrication of structural, large-scale deployable metamaterials.
}

\maketitle

\section*{Introduction}\label{Section:Introduction}
Origami, traditionally regarded as an art form, has recently attracted considerable attention for engineering applications \cite{RN676}.
Origami facilitates the fabrication of diverse functional shapes from flat sheets, including curved surfaces~\cite{RN16,RN959,RN956}.
Moreover, origami kinematics has been applied to deployable space structures \cite{RN368}, shelters \cite{RN56,RN471}, robotic systems \cite{RN482,RN906}, and self-folding structures \cite{RN910,RN480,RN53}.
In addition to planar origami, origami tubes have emerged as a prominent research topic and have been utilized in medical stents \cite{RN244}, energy-absorbing structures \cite{RN94}, and deployable flat surfaces \cite{RN911}.
Furthermore, origami tubes exhibit unique mechanical properties, including topological mechanics \cite{RN326,RN774}, decoupled kinematic and elastic modes \cite{RN13}, and unusual elastic wave modes \cite{RN62,RN63,RN713}.
Tessellations of origami tubes further enable advanced functionalities such as mode separation \cite{RN955}, bidirectional folding \cite{RN11}, auxetic behavior \cite{RN60}, and enhanced energy absorption \cite{RN105,RN258}.

While these macroscopic properties rely on three-dimensional tessellations of tubes, scaling these architectures for practical engineering applications requires origami metamaterials to be fabricated from thick panels to ensure sufficient load-bearing capacity.
To address interference caused by material thickness in manifold structures, such as planar origami and isolated origami tubes, various thickness-accommodation methods have been proposed \cite{RN225}, including hinge offsetting \cite{RN264,RN194} and panel trimming \cite{RN264}.
However, conventional thickness-accommodation techniques are generally difficult to apply to non-manifold geometries, where three or more thick panels converge at a single edge.
Therefore, conventional fabrication methods have typically sacrificed ideal rigid folding by introducing compliant panels \cite{RN911}, wide hinges \cite{RN15}, or sliding mechanisms \cite{RN798}, which often hinder smooth deployment and reduce structural predictability.
Although the coupling of thick Miura-ori tubes via creases using scissor-like linkages \cite{RN722} has been proposed to overcome these limitations, a practical fabrication strategy has not yet been established because the panels were assembled using adhesive tape, limiting structural strength and assembly efficiency.

\begin{figure}[H]
\includegraphics[clip, width=0.9\linewidth]{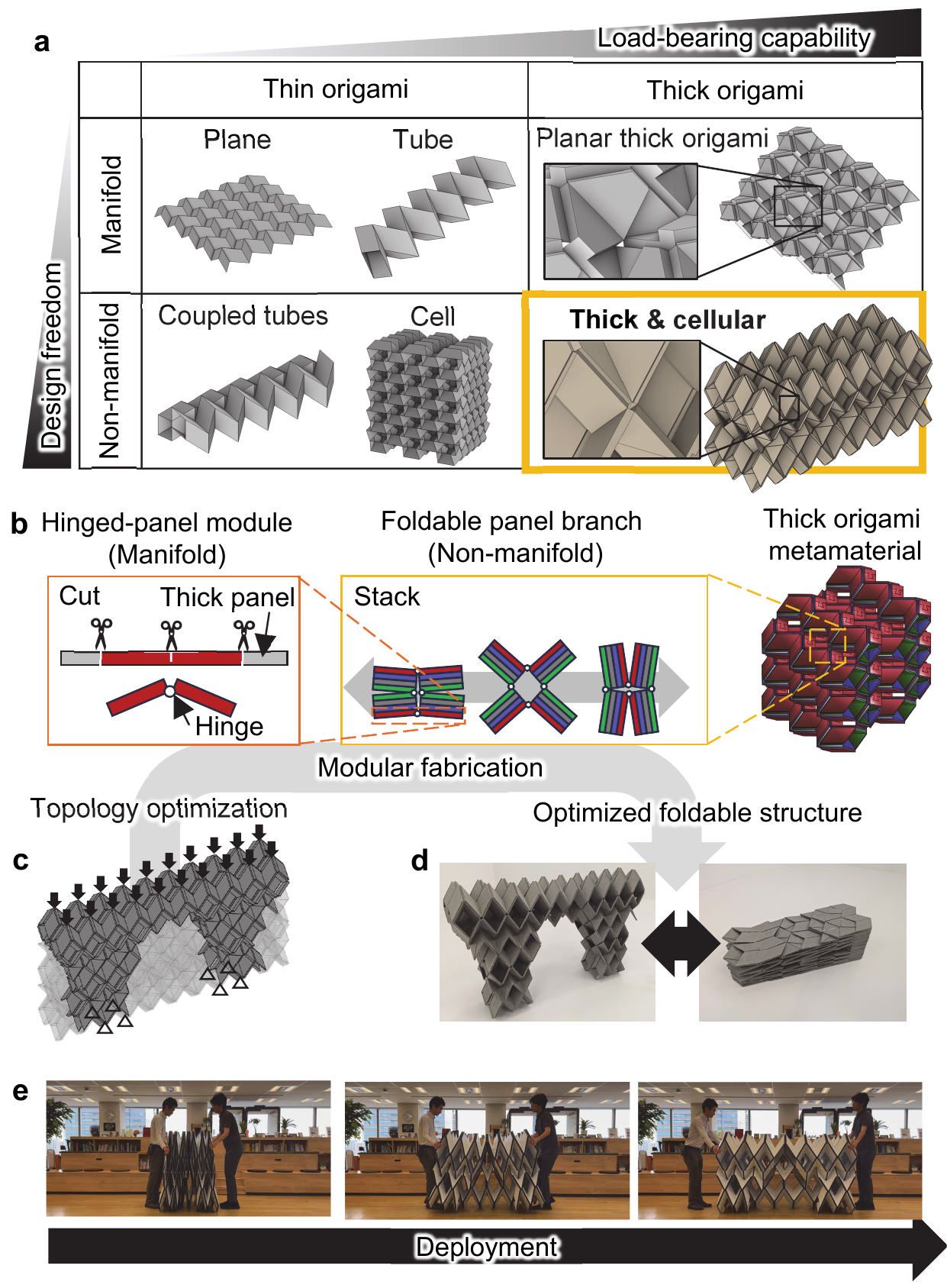}
\caption{\label{fig:Concept}
\textbf{Concept for modular design and fabrication of rigid-foldable thick origami metamaterial.}\\ 
\textbf{a} Classification of rigid-foldable origami in terms of thickness and manifold. This study aims to fabricate thick and non-manifold rigid-foldable origami metamaterials to achieve high design freedom and load-bearing capability.
\textbf{b} Schematic of modular fabrication of thick origami metamaterials.
The modular hinged panels are fabricated by cutting thick panels, which are then assembled to form origami metamaterials.
Kinematic motion of the non-manifold panel branch is maintained by the scissor-like linkages formed by stacked hinged panels, achieving a continuous fabrication process from the cutting of a planar plate to rigid-foldable, thick, and non-manifold origami. 
\textbf{c} Modular fabrication enables the optimized layout of panel distributions obtained by topology optimization.
\textbf{d} Modular hinged panels are placed based on the optimized panel distribution.
\textbf{e} Large-scale construction of thick rigid-foldable origami metamaterials.}

\end{figure}

This study proposes a modular fabrication method for rigid-foldable thick origami metamaterials that enables high design freedom and load-bearing capability (Fig.~\ref{fig:Concept}\textbf{a}), in which modules incorporating living hinges are rigidly joined, thereby improving both assembly efficiency and structural integrity.
The key concept of our modular fabrication approach is to decompose a non-manifold junction into a combination of a hinged-panel module (manifold level) and a foldable panel branch (non-manifold level), as shown in Fig.~\ref{fig:Concept}\textbf{b}.
By stacking these modularized hinges along the thickness direction, the proposed method accommodates the hinge placements required for kinematic compatibility while avoiding geometric interference.
To our knowledge, this approach is the first fabrication strategy for thick-panel origami metamaterials that remain rigid-foldable.
Furthermore, because these structures can be fabricated through simple cutting and lamination of thick panels, they can be readily produced using widely available digital fabrication tools such as cutting machines \cite{RN471}.

Beyond solving the fabrication challenge, this modular representation provides a discrete design space with exceptionally high geometric expressiveness.
Because the cellular structure is represented as a combination of discrete modular components, it enables the computational tailoring of macroscopic structural properties.
Leveraging this feature, we formulate a graph-based topology optimization framework to identify lightweight structures with minimized structural compliance (Fig.~\ref{fig:Concept}\textbf{c}).
Unlike traditional origami optimization, which is often limited to geometric shape tuning, our modular graph representation enables the optimized layout of panels while enforcing global folding connectivity.
In this framework, the design variables are filtered through a graph representation in which nodes represent hinged-panel modules and edges represent panel connectivity.
By optimizing the spatial layout of these modules under external loads while promoting the folding connectivity required for deployability, we can derive highly efficient, functional, and manufacturable structures, as shown in Fig.~\ref{fig:Concept}\textbf{d}, which can be constructed using the modular fabrication framework on a large scale, as shown in Fig.~\ref{fig:Concept}\textbf{e}.

This paper demonstrates that rigid-foldable origami metamaterials can be simultaneously manufacturable, mechanically efficient, and computationally optimizable, opening a new avenue toward structurally functional deployable metamaterials.

\section*{Results}\label{Section:Results}

\subsection*{Rigid-foldable thick origami metamaterials with non-manifold connectivity}\label{Subsection:Non-manifold hinges}
\begin{figure}[H]
\includegraphics[clip, width=0.9\linewidth]{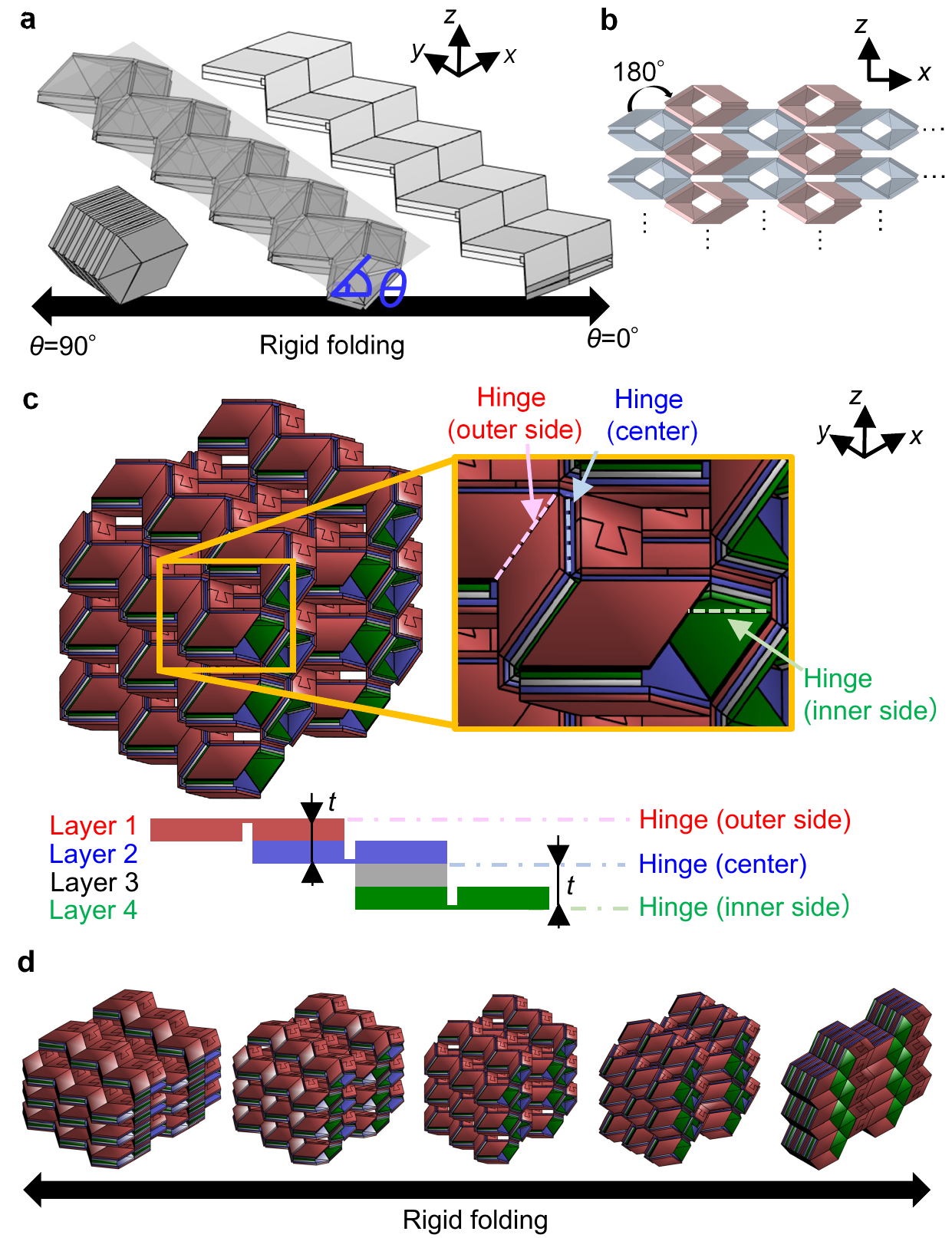}
\caption{\label{fig:Non-manifold}
\textbf{Modular hinged panels for rigid-foldable, non-manifold and thick origami metamaterials}\\
\textbf{a} Rigid-foldable thick origami tubes. 
The angle $\theta$ characterizes the extension of the origami tubes.
\textbf{b} Spatial tessellation of thick origami tubes for cellular assembly by repeated coupling of relative 180$^\circ$ rotated tubes.
\textbf{c} Separation of hinged panels for the construction of a thick origami metamaterial to shift hinge locations along the panel thickness for valid one-degree-of-freedom kinematics.
\textbf{d} Rigid-foldable motion of the assembled thick origami metamaterials, which can transition seamlessly between two flat-folded states.
}
\end{figure}

The geometric foundation of the proposed design originates from the rigid-folding kinematics of a thick Miura-ori tube (Fig.~\ref{fig:Non-manifold}\textbf{a})~\cite{RN191}.
The configuration and continuous deployment of the single tube are governed by the folding angle $\theta$ (the half dihedral angle between adjacent panels).
Under the rigid-origami assumptions, the tube possesses a one-degree-of-freedom (1-DOF) kinematic motion, allowing it to transition seamlessly and reversibly from a flat state ($\theta = 0^\circ$) to another flat state ($\theta = 90^\circ$). 
To construct a spatial cellular structure, the thick Miura-ori tubes are periodically tessellated by arranging them via relative 180$^\circ$ rotations along their longitudinal axes (Fig.~\ref{fig:Non-manifold}\textbf{b})~\cite{RN722}.
This tessellation preserves the 1-DOF folding kinematics of the original individual tubes.

To resolve the material interference and preserve the ideal rigid-folding motion within the thick-panel cellular structure, we systematically distribute the hinge axes through the panel thickness by decomposing each thick panel into a multi-layered panel module (Fig.~\ref{fig:Non-manifold}\textbf{c}). 
The compatibility of the folding requires the hinges to be positioned at three distinct out-of-plane locations: the outer surface, the mid-thickness region, and the inner surface of the tubular wall. 

Specifically, as shown in the layered cross-section in Fig.~\ref{fig:Non-manifold}\textbf{c}, the outer valley hinge is located on the outer boundary of Layer~1.
To accommodate the spatial clearances necessitated by adjacent folding panels, the mountain hinge on Layer~2 is geometrically shifted inward by a distance corresponding to the subpanel thickness.
Layer~3 contains no crease, functioning as a spacer layer that prevents physical interference and constraint.
Finally, Layer~4 incorporates the inner mountain hinge whose axis is positioned on the inner surface of the tube wall, completing the spatial shift relative to the preceding layers.

This engineered layered arrangement ensures that these structural offsets are perfectly absorbed within the internal space of the module.
Consequently, the local geometric compatibility at the non-manifold junction is entirely satisfied without sacrificing rigid-body panels or introducing loose joint clearances. 

As a result, unlike structures relying on self-folding \cite{RN910,RN480,RN53} or additive manufacturing \cite{RN15,RN456}, the assembled structural system successfully scales up from individual components into a coherent, three-dimensional rigid-foldable cellular metamaterial (Fig.~\ref{fig:Non-manifold}\textbf{d}).
As demonstrated in the continuous deformation sequence, the entire thick cellular assembly can undergo a highly synchronized, unconstrained 1-DOF rigid-folding transformation throughout the full range of $\theta$.
Consequently, the proposed layered hinge architecture establishes a manufacturable design principle for thick non-manifold origami metamaterials, enabling rigid folding without sacrificing structural thickness.

\subsection*{Fabrication by modular hinged panels}\label{Subsection:Assembly}
\begin{figure}[H]
\includegraphics[clip, width=0.9\linewidth]{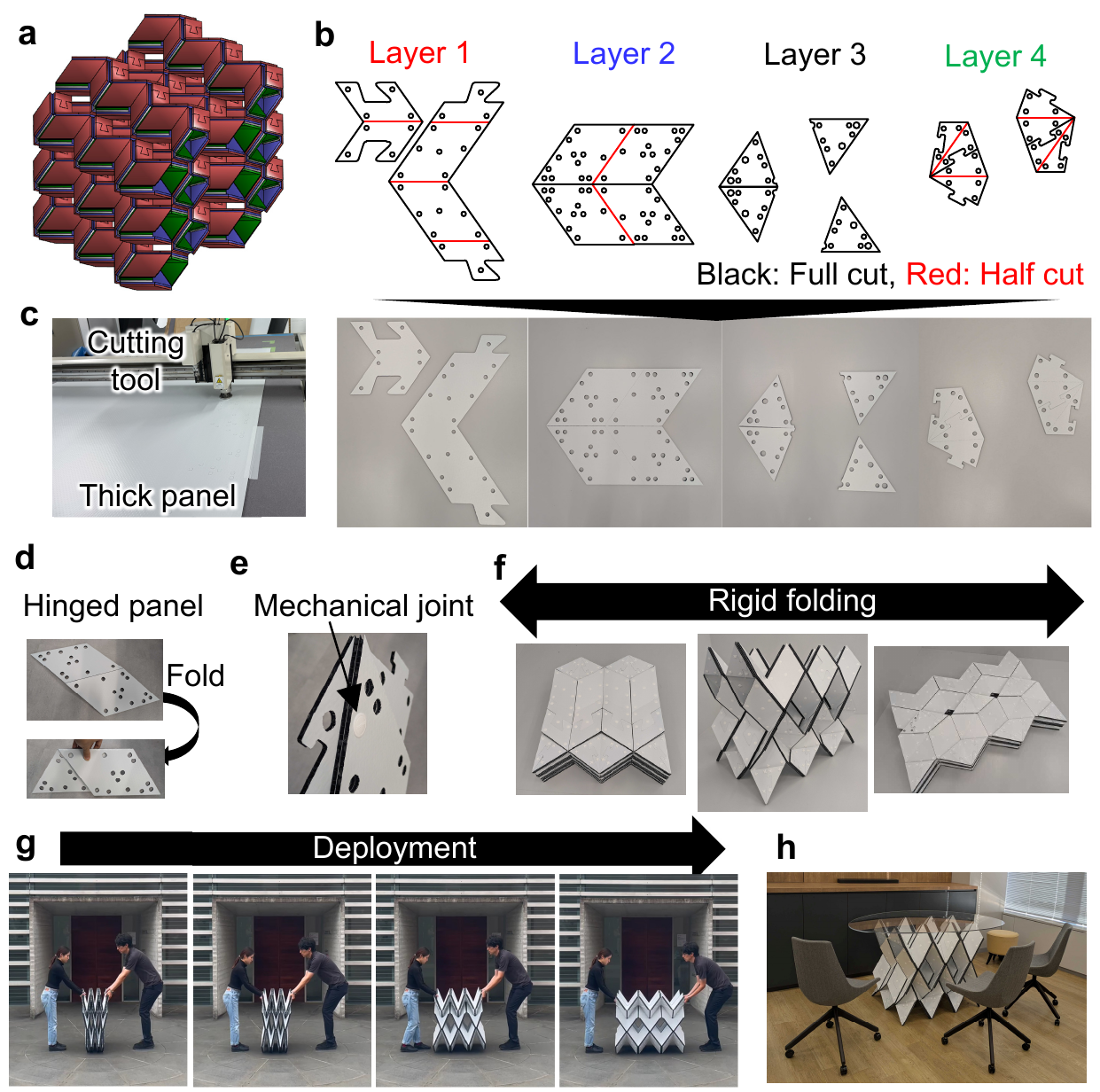}
\caption{\label{fig:Fabrication}
\textbf{Modular fabrication of rigid-foldable thick origami metamaterial}\\
\textbf{a} Thick origami metamaterial consisting of 4 layered hinged panels.
\textbf{b} Cutting line of the hinged thick panels. The black and red lines indicate the full and half cuts, respectively. The holes on the panels are used for mechanical joints.
\textbf{c} Thick panels are cut out by a cutting plotter.
\textbf{d} Foldability of the hinged thick panel.
\textbf{e} The hinged thick panels are stacked via mechanical joints.
\textbf{f} The assembled thick origami metamaterial exhibits rigid-foldable motion.
\textbf{g} Deployment of the assembled origami metamaterial via one-degree-of-freedom motion.
\textbf{h} The deployed origami metamaterial holds an 11~kg tabletop and functions as a deployable load-bearing structure.
}
\end{figure}

For the construction of the origami metamaterials by stacking modular hinged panels (Fig.~\ref{fig:Fabrication}\textbf{a}) based on the stacked layers introduced in Fig.~\ref{fig:Non-manifold}, the cutting patterns of the panels for Layers $1 \sim 4$ are processed to form the designed cutting patterns (Fig.~\ref{fig:Fabrication}\textbf{b}) using a flatbed cutting plotter (Fig.~\ref{fig:Fabrication}\textbf{c}).
To preserve rigid-folding compatibility, the hinge axes are distributed across the panel thickness by stacking modular hinged panels.
This hinge configuration is crucial for geometric compatibility, which effectively prevents material interference during the folding of the thick panels, thereby ensuring valid kinematic motion.

The remaining skin material at the half-cut lines functions as a compliant living hinge, eliminating the need for complex mechanical hinges at each fold line.
As demonstrated in Fig.~\ref{fig:Fabrication}\textbf{d}, these half-cut lines allow the panels to fold flexibly, transforming the flat sheets into functional hinged modules. 

The modular hinged panels are subsequently stacked and assembled using mechanical fasteners inserted through pre-drilled positioning holes (Fig.~\ref{fig:Fabrication}\textbf{e}).
This strategic stacking process enables the branches of more than three panels (non-manifold origami) to deform via scissor-like linkages.
This transformation yields a rigid-foldable, thick origami metamaterial capable of transitioning seamlessly between two fully flat states through a synchronized rigid-foldable motion (Fig.~\ref{fig:Fabrication}\textbf{f}).

Figure~\ref{fig:Fabrication}\textbf{g} illustrates the continuous deployment motion of the fabricated thick origami metamaterial.
Owing to its 1-DOF mechanism, the structure can be effortlessly deployed from its compact state simply by pulling its opposing left and right ends. 

To validate the engineering feasibility and structural integrity of the system for practical structural applications (e.g., deployable architectural furniture), the deployed metamaterial was subjected to a static load test supporting a tabletop weighing approximately 11~kg (Fig.~\ref{fig:Fabrication}\textbf{h}).
While the metamaterial remains highly flexible during the deployment phase, it acquires high load-bearing capacity upon reaching its deployed state.

In summary, the proposed modular fabrication methodology offers two distinct features for metamaterial design:
\begin{enumerate}
    \item Kinematics and stiffness: The thick origami metamaterial successfully decouples folding flexibility during deployment from load-bearing capability in the operational state.
    \item Scalability and manufacturability: The partitionable and modular design enables the large-scale construction of extensive deployable structures via the assembly of smaller, separate panels.
    This effectively circumvents the physical working-area limitations imposed by standard fabrication equipment, such as cutting plotters.
\end{enumerate}

Consequently, compared with conventional origami metamaterials that rely on thin, continuous sheets, the proposed approach significantly extends the structural scalability of thick origami through the synergy of enhanced load-carrying capacity and a modular fabrication framework.

\newpage
\subsection*{Topology optimization of origami metamaterials}\label{Subsection:TO}
\begin{figure}[H]
\includegraphics[clip, width=0.9\linewidth]{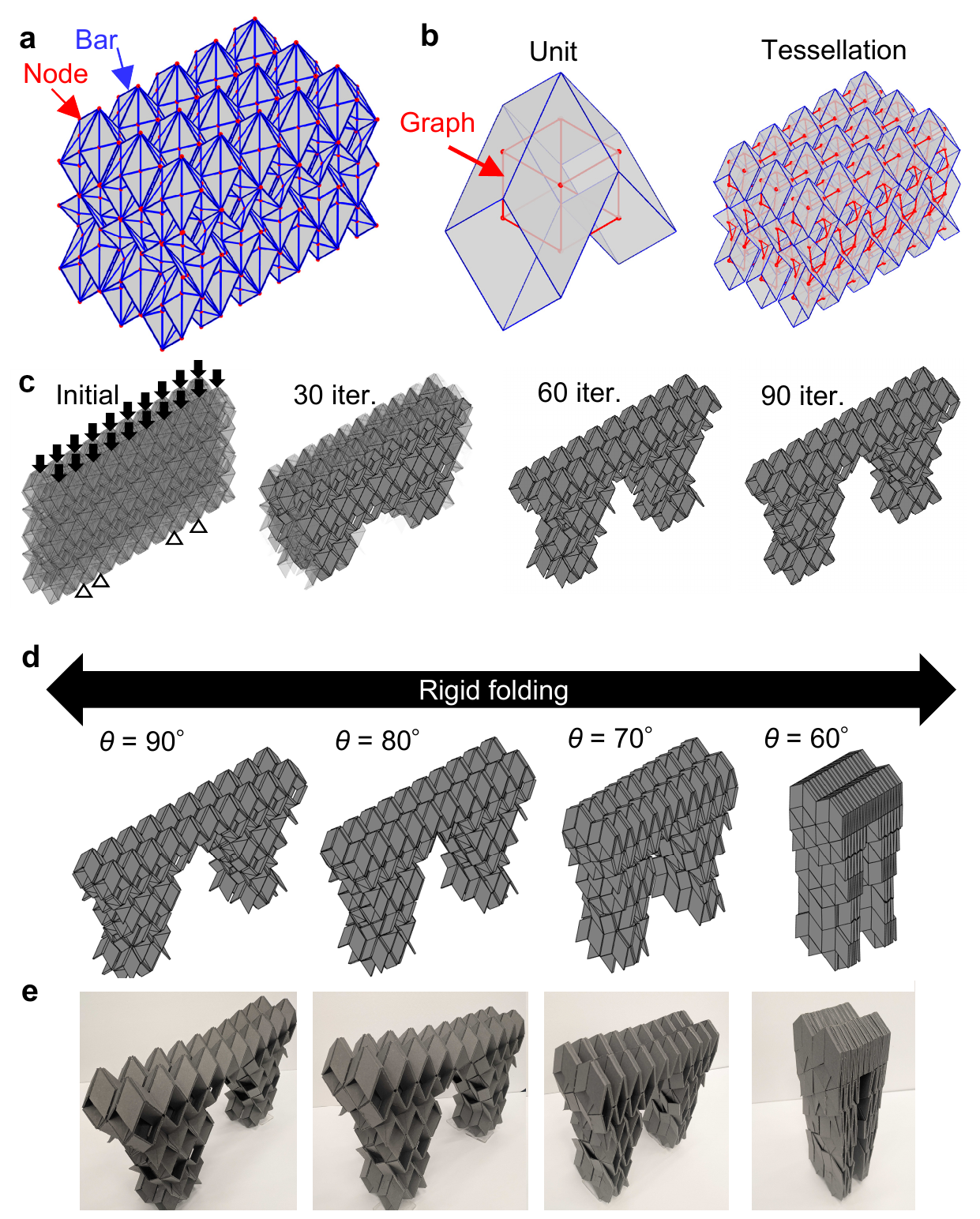}
\caption{\label{fig:TO}
\textbf{Topology optimization of thick origami tessellation.}\\
\textbf{a} Physical representation of the origami tessellation by a bar-and-hinge model. The design variables are assigned to the panels and characterize the stiffness of the bars and rotational springs on the diagonal bars.
\textbf{b} Graph structure of panel connectivity, where the density field is filtered using the shortest-path distance on the graph.
\textbf{c} Boundary conditions and density history during the topology optimization process.
\textbf{d} Rigid-folding kinematics of the optimized origami tessellation.
\textbf{e} Fabricated physical prototype validating the kinematics generated by the modular fabrication approach.
}
\end{figure}

The topology optimization was performed on a ground structure with a density filter defined by a graph structure constructed from tessellated thick Miura-ori tube units.
As shown in Fig.~\ref{fig:TO}\textbf{a}, each panel was represented using bars and hinges, which model the panel deformation \cite{RN196,RN17}.
The ground structure was generated by periodically tessellating the tubular unit cell shown in Fig.~\ref{fig:Non-manifold}, resulting in a design space that preserves the rigid-foldable kinematics of the underlying origami metamaterials.

Figure~\ref{fig:TO}\textbf{b} illustrates the graph representation of panel connectivity in the Miura-ori tube, alongside the resulting tessellated ground structure.
The graph-based formulation enhances connectivity during the optimization process, thereby promoting the formation of continuous tubular networks while suppressing isolated panels.

A compressive load was applied to the top surface of the structure, whereas selected nodes on the bottom surface were fixed, as indicated in Fig.~\ref{fig:TO}\textbf{c}.
Starting from a uniform density distribution, the optimization progressively removed mechanically inefficient panels while promoting the connectivity characteristic of the underlying origami tessellation.
The evolution of the material distribution is shown after 30, 60, and 90 optimization iterations. 
As the optimization proceeds, the structure gradually converges toward a branched load-carrying topology connecting the loading and support regions.

The optimized structure retains the kinematic compatibility of the original tessellation and can therefore be deployed through rigid folding.
Figure~\ref{fig:TO}\textbf{d} shows the deformation sequence of the optimized design during folding, where the folding angle is varied from $\theta = 90^\circ$ to $60^\circ$.
Despite the removal of panels, the optimized structure preserves a continuous folding motion without geometric interference or loss of connectivity.

To validate the manufacturability of the optimized design, a physical prototype was fabricated using the layered hinged-panel construction.
The fabricated structures at different folding angles are presented in Fig.~\ref{fig:TO}\textbf{e}.
The prototype exhibits the same deployment behavior predicted by the geometric model, confirming that the proposed graph-based topology optimization framework can generate mechanically efficient structures while maintaining rigid-foldable deployability.

This computational framework extends topology optimization to deployable origami metamaterials by simultaneously considering mechanical performance and folding compatibility.

\newpage

\subsection*{Large-scale construction}\label{Subsection:Large}
One of the primary advantages of the proposed modular fabrication strategy is its ability to overcome the size limitations imposed by conventional sheet-based origami manufacturing and additive manufacturing.
Because the structure is assembled from independent hinged-panel modules rather than fabricated from a single continuous sheet, the overall dimensions are not constrained by the working area of the cutting equipment.
Consequently, large-scale deployable structures can be realized through the assembly of a large number of modular components.

\begin{figure}[H]
\includegraphics[clip, width=0.9\linewidth]{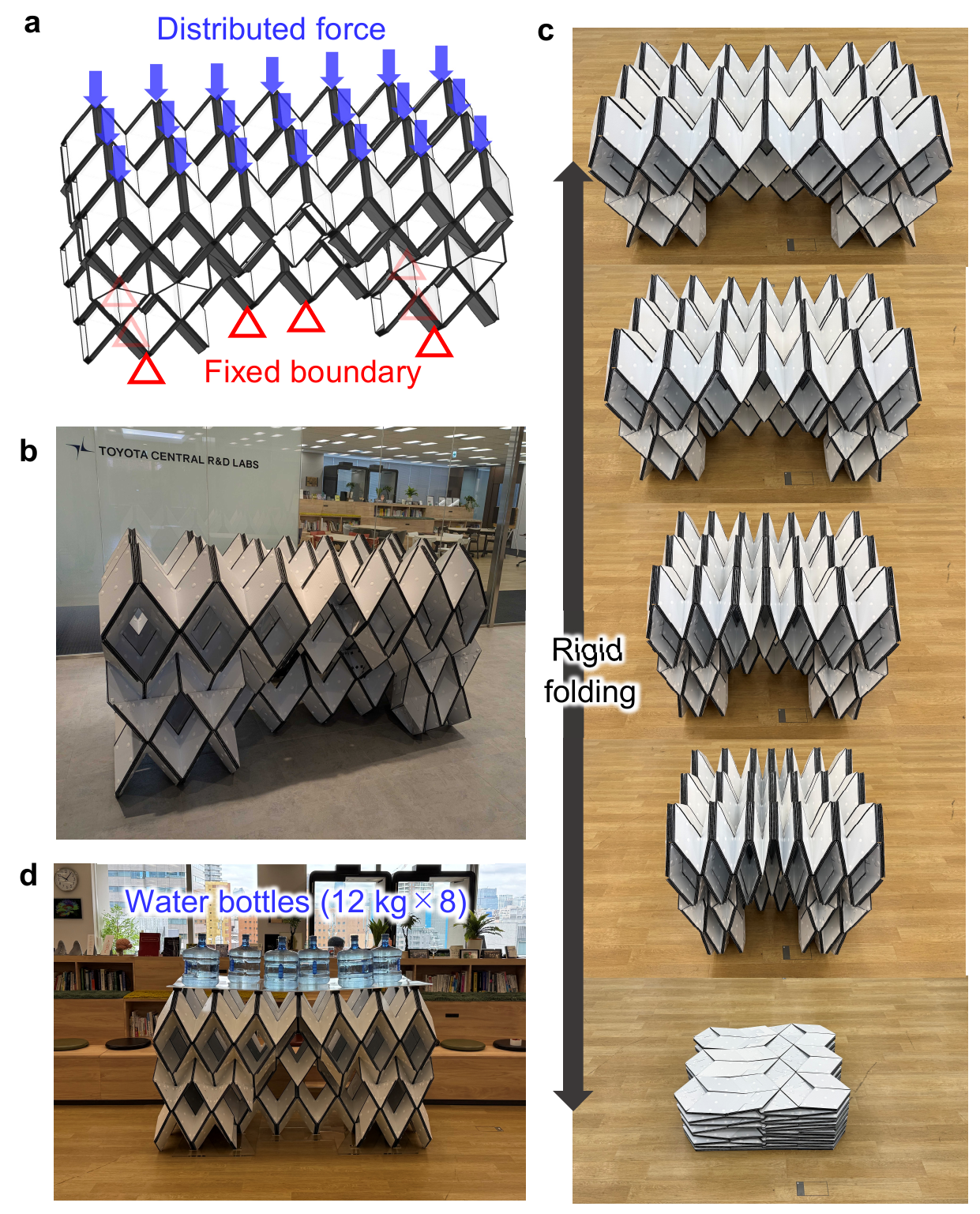}
\caption{\label{fig:LargeScaleFab}
\textbf{Large-scale design and fabrication of thick origami metamaterial.}\\
\textbf{a} Loading and boundary conditions used for structural evaluation. A distributed vertical load is applied to the upper surface of the metamaterial, while selected nodes on the bottom surface are fixed.
\textbf{b} Fabricated thick-origami metamaterial in the deployed configuration.
\textbf{c} Rigid-folding sequence of the fabricated structure. The metamaterial undergoes a continuous one-degree-of-freedom deployment motion between compact and deployed configurations while preserving geometric compatibility.
\textbf{d} High load-bearing demonstration. The deployed metamaterial supports eight water bottles (approximately 96~kg in total), demonstrating its ability to sustain substantial compressive loads.
}
\end{figure}

To demonstrate this scalability, a large thick-origami metamaterial consisting of tessellated Miura-ori tubes was designed by minimizing the compliance under distributed forces on the top surface, as shown in Fig.~\ref{fig:LargeScaleFab}\textbf{a}.
Based on the panel distribution, the thick origami metamaterial was fabricated, as shown in Fig.~\ref{fig:LargeScaleFab}\textbf{b}.
The structure was assembled from modular hinged panels, which enabled the assembled metamaterial to maintain the rigid-foldable kinematics of the original unit cell despite its large size.

The deployability of the fabricated structure was subsequently evaluated, as shown in Fig.~\ref{fig:LargeScaleFab}\textbf{c}, which illustrates the rigid-folding sequence from the deployed state to the compact state.
The metamaterial exhibits a 1-DOF folding motion without mechanical interference between adjacent modules.
Owing to the modular hinge arrangement introduced through the layered panel design, the structure preserves geometric compatibility throughout the folding process and can be repeatedly transformed between compact and deployed configurations.

As a proof-of-concept demonstration for the high load-bearing capability of thick origami metamaterials, eight water bottles (approximately 96~kg in total) were placed on the metamaterial (Fig.~\ref{fig:LargeScaleFab}\textbf{d}).
No visible structural failure or collapse was observed during the test, confirming that the deployed metamaterial can sustain substantial compressive loads.

While the structure remains highly compliant during deployment, the stacked panels progressively establish direct load-transfer paths after deployment.
This behavior enables the coexistence of compact deployability and structural load-bearing capability within a single structure on a meter scale.
The combination of modular fabrication and rigid-foldable kinematics, therefore, enables deployable structures that are simultaneously scalable, load-bearing, and manufacturable.

\section*{Discussion}\label{Section:Discussion}
The modular design, optimization, and fabrication framework developed in this study establishes a manufacturing and design framework for functional, thick-panel metamaterials.
A key mechanical insight of this approach is the realization of kinematic and structural functionality.
During the deployment phase, the metamaterial behaves as a highly compliant mechanism, owing to the localized out-of-plane rotation of the living hinges formed by the skin material.
However, upon reaching the deployed state, the thick panels efficiently transfer mechanical loads within the metamaterial.
Consequently, the structure transitions from a flexible deployable mechanism into a robust load-bearing assembly capable of supporting static loads.

Furthermore, this modular framework fundamentally improves manufacturing scalability of deployable structural systems.
Our approach circumvents the size limitation by allowing large-scale, extensive deployable cellular metamaterials to be partitioned and constructed from smaller, segmented panel assemblies (Fig.~\ref{fig:LargeScaleFab}).
This modular approach, paired with our graph-based topology optimization framework, enables the algorithmic removal of structurally inefficient panel pathways.
The optimization not only achieves lightweight targets but also reduces material usage, all while preserving global kinematic compatibility.

In conclusion, this work bridges the gap between kinematic motions in large-scale structures and load-bearing structural design.
The synthesis of modular fabrication, graph-based topology optimization, and segmented structural assembly provides a scalable strategy for functional deployable structures.
Beyond deployable structures, such structures scale up adaptive metamaterials based on lattice transformations for controlling phenomena such as acoustic~\cite{RN957,RN773} and electromagnetic~\cite{RN958,RN960} waves.
Therefore, the fundamental concepts introduced here can be readily extended to a wide spectrum of applications across scales, including deployable bridges, rapidly deployable disaster-relief infrastructure and reconfigurable sound- or electromagnetic-wave-absorbing structures.

\section*{Methods}\label{Section:Methods}
\subsection*{Fabrication of thick origami metamaterials}
Large-scale thick origami metamaterials shown in Figs.~\ref{fig:Fabrication} and \ref{fig:LargeScaleFab} were fabricated using sandwich panels with a nominal thickness of 5~mm (PDPPZ-090, Kawakami Sangyo Co., Ltd.).
The structured, hollow polypropylene (PP) core of these sandwich panels provides high out-of-plane bending stiffness while remaining lightweight. 

To achieve reliable and reproducible kinematics, the cutting depth of the flatbed cutting plotter (Fig.~\ref{fig:Fabrication}\textbf{c}) was regulated to remain within the surface layer of the sandwich panel to properly form the compliant living hinges.
For the layer assembly, 10-mm long mechanical fasteners (Panlock $\phi$15$\times$L10, Kunimori Chemical Industry Co., Ltd.) were used for Layers 1 and 2 through the pre-drilled holes shown in Fig.~\ref{fig:Fabrication}\textbf{d}.
Similarly, 15-mm long fasteners (Panlock $\phi$15$\times$L15, Kunimori Chemical Industry Co., Ltd.) were employed to secure Layers 2 to 4.

The small prototype shown in Fig.~\ref{fig:TO}\textbf{e} was fabricated from 1-mm-thick paperboard and cut using a cutting plotter.
Double-sided adhesive tape (No.~5000NS, Nitto Denko Corporation) was used to facilitate the assembly of the origami cellular structure.

\subsection*{Formulation of topology optimization}\label{Subsection:TOformulation}
The proposed topology optimization framework is developed for deployable thick-origami metamaterials constructed from tessellated Miura-ori tube units. The structural response is evaluated using a bar--hinge model, while the material distribution is filtered through a graph structure.

A design variable $\phi \in [-1,1]$ is assigned to each panel.
To preserve the connectivity of the tessellated tube network, a graph-based density filter is introduced:

\begin{equation}
\tilde{\phi}_e
=
\frac{\sum_j w(d_{ej}) \phi_j}
{\sum_j w(d_{ej})},
\label{eq:graphfilter}
\end{equation}
where $d_{ej}$ denotes the shortest-path distance between panels on the connectivity graph and the weighting function is defined as:
\begin{equation}
	w(d_{ej})=
	\begin{cases}
			\frac{R_\mathrm{min}-d_{ej}}{R_{\mathrm{min}}} & \text{if $d_{ej}<R_\mathrm{min}$,} \\
			0       & \text{otherwise.}
		\end{cases}
\end{equation}
Here, $R_{\mathrm{min}}$ is the filter radius.

A Heaviside projection is subsequently applied to promote binary designs,

\begin{equation}
\rho_e
=
\frac{\tanh(\beta \tilde{\phi}_e)}
     {2\tanh(\beta)}
+\frac{1}{2},
\label{eq:heaviside}
\end{equation}
where $\beta$ controls the projection sharpness.

The panel thickness is interpolated using the SIMP \cite{RN592},

\begin{equation}
t_e = {\rho}_e^{p} t_0 ,
\label{eq:simp}
\end{equation}
where $t_0$ is the thickness of the solid panel and $p$ is the penalization factor.
The optimization problem is formulated as
\begin{align*}
\min_{\boldsymbol{\phi} \in [-1,1]} \quad & c + f_{\mathrm{quad}} + f_{\mathrm{conn}} \\
\text{subject to:}\ & \mathbf{K}\mathbf{U} = \mathbf{F} \\
& \sum_{e=1}^{N} \rho_e / N < V_f \\
& 0 < \rho^{\mathrm{min}} \le \rho_e \le 1
\end{align*}
where $V_f$ represents the volume fraction of the panels on the ground structure, $N$ denotes the total number of panels within the ground structure, and $\boldsymbol{\rho}$ is the panel densities.

The first term of the objective function $c$ represents the structural compliance. Utilizing the global stiffness matrix $\mathbf{K}$ and the displacement vector $\mathbf{U}$, it is expressed as:
\begin{equation}
c({\boldsymbol{\rho}}) = \mathbf{U}^\top \mathbf{K}({\boldsymbol{\rho}}) \mathbf{U} = \sum_{e=1}^{N} \mathbf{u}_{e}^\top \mathbf{k}_{Se} \mathbf{u}_{e} + \sum_{e=1}^{N} \mathbf{u}_{e}^\top \mathbf{k}_{Be} \mathbf{u}_{e},
\label{eq: Sec3_Complince}
\end{equation}
where $\mathbf{u}_{e}$ denotes the elemental translational displacement vector.
Based on the stiffness formulation of the bar-and-hinge model \cite{RN17}, the local stiffness matrices associated with in-plane deformation ($\mathbf{k}_{Se}$) and out-of-plane deformation ($\mathbf{k}_{Be}$), parameterized by the panel density ${\rho}_e$, are given by:
\begin{equation}
\mathbf{k}_{Se} = {\rho}_e^p \mathbf{k}_{S0}, 
\end{equation}
and
\begin{equation}
\mathbf{k}_{Be} = {\rho}_e^{(8/3)p} \mathbf{k}_{B0},
\end{equation}
respectively.
Here, $\mathbf{k}_{S0}$ and $\mathbf{k}_{B0}$ represent the baseline stiffness matrices evaluated at a reference thickness $t = t_0$.

The second term, $f_{\mathrm{quad}}$ promotes length-four closed walks in the weighted connectivity graph and thereby favors local connectivity patterns associated with four-panel tube configurations:

\begin{equation}
f_{\mathrm{quad}} = -\lambda_{\mathrm{quad}} \cdot \mathrm{Tr}(\mathbf{A}^4(\mathbf{q})),
\label{eq:Penalty1}
\end{equation}
where $\lambda_{\mathrm{quad}}$ is a penalty coefficient.
Given an adjacency matrix $\mathbf{A}$, the trace property $\mathrm{Tr}(\mathbf{A}^r) = \sum_{i} (\mathbf{A}^r)_{ii}$ represents the number of closed walks of length $r$ that start and end at the same node \cite{RN905}.
Accordingly, to promote length-four closed walks in the connectivity graph, we set $r = 4$ to enforce this specific connectivity.
Here, $\mathbf{q}$ denotes the weight vector of the connectivity graph composed of adjacent nodes $i$ and $j$; its components are defined based on the panel densities as $q_{ij} = ({\rho}_i + {\rho}_j)/2$.

The third term, $f_{\mathrm{conn}}$ penalizes panels with insufficient local connectivity and is defined as:
\begin{equation}
f_{\mathrm{conn}}({\boldsymbol{\rho}}) = \lambda_{\mathrm{conn}}\sum_{i=1}^{N} {\rho}_i \cdot \left[ \max\left(0, d_{\mathrm{min}} - \sum_{j=1}^{N} \mathbf{A}_{ij} {\rho}_j \right) \right]^2
\label{eq:f_conn}
\end{equation}
where $\lambda_{\mathrm{conn}}$ is a penalty coefficient. This term penalizes graph topologies whose degree of connectivity falls below a specified minimum threshold $d_{\mathrm{min}}$.

In this study, the penalization factor was set to $p=3$, the filter radius to $R_\mathrm{min}=4.5$, and the volume fraction to $V_f=0.6$ for Fig.~\ref{fig:TO}\textbf{c} and $V_f=0.85$ for Fig.~\ref{fig:LargeScaleFab}\textbf{a}. 
The parameter $\beta$ was initialized to 1 and increased by $20\%$ after each optimization iteration until exceeding 30,000, after which it was kept fixed.
The coefficients in Eqs.~\eqref{eq:Penalty1} and \eqref{eq:f_conn}, namely $\lambda_{\mathrm{quad}}$ and $\lambda_{\mathrm{conn}}$, were determined at each step so that the compliance $c$ and each penalty term maintained an equivalent scale.
The Method of Moving Asymptotes (MMA) \cite{RN648} was employed for the optimization process.

\section*{Data availability}\label{Data availability}
Data sets generated during the current study are available from the corresponding author on reasonable request.

\section*{Author contributions}\label{Section:Author contributions}
S.T. and T.T. conceptualized this study.
S.T. wrote the manuscript, designed and fabricated the structures, and implemented the numerical simulation.
H.K., S.A. and T.N. supported the fabrication of metamaterials.
M.T., A.K., and T.N. advised on the numerical modeling and structural optimization.
T.T. contributed to the theoretical foundation of the kinematics and geometric compatibility.
All authors contributed to the discussion of the results and manuscript preparation.

\section*{Competing interests}\label{Competing interests}
The authors declare no competing interests.

\bibliography{ThickOrigami}

\section*{Funding}\label{Funding}
This research received no specific grant from any funding agency in the public, commercial, or not-for-profit sectors.

\end{document}